\documentclass[aps,prl,a4,twocolumn,superscriptaddress,nofootinbib,showpacs,showkeys,preprintnumbers]{revtex4-1}
\usepackage{amsfonts}
\usepackage{mathrsfs}
\usepackage{amssymb}
\usepackage{amsmath}
\usepackage{graphicx,epsfig}
\usepackage{pstricks}
\usepackage{slashed}
\usepackage{dcolumn}%
\usepackage[margin=9pt]{subfig}
\usepackage{color,ulem}
\usepackage[utf8]{inputenc}
\usepackage{mathtools}

\def\bs{\begin{small}}
\def\es{\end{small}}
\def\be{\begin{equation}}
\def\ee{\end{equation}}
\def\bea{\begin{eqnarray}}
\def\eea{\end{eqnarray}}
\def\bean{\begin{eqnarray*}}
\def\eean{\end{eqnarray*}}
\def\bary{\begin{array}}
\def\eary{\end{array}}
\def\bit{\begin{itemize}}
\def\eit{\end{itemize}}

\def\su5u1{SU(5) \times U(1)}
\def\fsu5u1{SU(5) \times U(1)'}
\def\so10{SO(10)}
\def\sq20{SO(10) \times SO(10)}

\def\bwt{\begin{widetext}}
\def\ewt{\end{widetext}}
\def\be{\begin{equation}}
\def\ee{\end{equation}}
\def\bea{\begin{eqnarray}}
\def\eea{\end{eqnarray}}
\def\bean{\begin{eqnarray*}}
\def\eean{\end{eqnarray*}}
\def\bary{\begin{array}}
\def\eary{\end{array}}
\def\bit{\begin{itemize}}
\def\eit{\end{itemize}}

\def\su5u1{SU(5) \times U(1)}
\def\fsu5u1{SU(5) \times U(1)'}
\def\so10{SO(10)}
\def\sq20{SO(10) \times SO(10)}

\def\Zp{Z^{\prime}}

\usepackage{amssymb}

\begin{document}

\setlength{\parskip}{0.1cm}

\preprint{HRI-RECAPP-2015-021}

\title{\large The 750 GeV Diphoton excess in a $U(1)$ hidden symmetry model}

\author{Kasinath Das}
\email{kasinathdas@hri.res.in}
\affiliation{Regional Centre for Accelerator-based Particle Physics, Harish-Chandra Research Institute,
Chhatnag Road, Jhusi, Allahabad 211019, India}
\author{Santosh Kumar Rai}
\email{skrai@hri.res.in}
\affiliation{Regional Centre for Accelerator-based Particle Physics, Harish-Chandra Research Institute,
Chhatnag Road, Jhusi, Allahabad 211019, India}

\begin{abstract}
Recent results from the experimental collaborations at LHC give  
hints of a resonance in the diphoton channel at an invariant mass of 750 GeV. 
We show that such a scalar resonance would be possible in an $U(1)$ extension 
of the SM where the extended symmetry is hidden and yet to be discovered. 
We explore the possibilities of accommodating this excess by introducing a minimal 
extension to the matter content and highlight the parameter space that can accommodate 
the observed diphoton resonance in the model. The model also predicts new interesting signals
that may be observed at the current LHC run.
\end{abstract}


\keywords{}

\maketitle


\section{Introduction}
Recent results from the ATLAS and CMS collaborations have shown the data 
from LHC run II with center of mass energy $\sqrt{s} = 13$ TeV \cite{lhc1,lhc2}. Interestingly,
the ATLAS data shows an excess in diphoton channel with 3.2 fb$^{-1}$ data giving about 
14 events (with selection efficiency 0.4 \cite{Franceschini:2015kwy}) at an invariant mass of $\sim$ 750 GeV. The 
local significance is slightly northward of $3.5\sigma$. On a lesser significance of about $2.6\sigma$,
a similar feature is exhibited by the CMS data with integrated luminosity
of 2.6 fb$^{-1}$, giving about 10  events, peaked at an invariant mass of 760 GeV.  The above rates with 
aforementioned efficiency corresponds to a rough order of magnitude cross section of $\sim 10$ fb for the 
$pp \to X \to \gamma \gamma$. 
Although this can be a mere fluctuation in the early observations of the data at the upgraded energy run of 
LHC, the fact that both the collaborations observe it makes it an intriguing prospect for new 
physics signals. This naturally has led to a plethora of ideas explaining the 
excess \cite{Harigaya:2015ezk}--\cite{Chakraborty:2015jvs}.

In this work we show that a simple extension to the SM gauge symmetry with a minimal set of new 
particles can easily accommodate the excess without invoking a large enough scale for new physics.
In addition the model predicts some interesting signals that could show up as more data is accumulated
in the run II of LHC. We consider an extra hidden $U(1)$ symmetry \cite{Grossmann:2010wm} in which 
all the SM particles are neutral.  Only new exotic quarks, and an electroweak (EW) singlet Higgs
boson can couple to this extra $U(1)$ gauge boson and the $U(1)$ symmetry is broken at the EW 
scale by the vacuum expectation value (VEV) of the EW singlet Higgs boson. In 
addition to this we extend the spectrum further by introducing an extra scalar which 
is a singlet under SM as well as the extra $U(1)$ symmetry \cite{Karabacak:2014nca}. We show that this 
scalar can be easily used to accommodate the observed diphoton excess with all  particles of the model 
having masses within the TeV scale. In addition, we highlight new exotic decay modes of the 
vector-like quark in the model that could give interesting signals at the LHC as well as a light 
sub-TeV $\Zp$ not constrained by existing experimental constraints. 
\section{Model}
The gauge symmetry in our model \cite{Grossmann:2010wm} is the usual standard model (SM): 
$SU(3)_C \times SU(2)_L \times U(1)_Y$ supplemented by an extra $U(1)$ symmetry, which 
we call $U(1)_X$.  We introduce two exotic quarks $xq_L$ and $xq_R$ which are color triplets but 
singlets under the $SU(2)_L$ gauge symmetry. They carry charge under the $U(1)_Y$ 
which decides whether they mix with the up-type or down-type SM quarks. We denote the gauge
boson for the $U(1)_X$ by $Z^\prime$. We introduce a complex Higgs field $S_1$ which 
acquires a VEV $v_1$ and breaks the $U(1)_X$. Therefore this scalar is a color and EW singlet, and 
has a charge $q^\prime$ under the $U(1)_X$. We also introduce a real scalar 
$S_2$ which is a singlet under $SU(3)_C \times SU(2)_L \times U(1)_Y \times U(1)_X$. 

The EW gauge interaction Lagrangian for the exotic
$xq$ quark is given by:
\begin{align}
\label{kinetic} \mathcal{L}&=
\overline{xq} \, i\slashed{\mathcal{D}} \, xq
\end{align}
where the covariant derivative is defined as
\begin{align}
\label{covariant}
\begin{split}
{\mathcal{D}}_{\mu} &= \partial_\mu - i \frac{g^\prime}{2} Y B_\mu - i g_{X} Y_{X} \Zp{_\mu},
\end{split}
\end{align}
and $Y_{X}$ is the charge of the matter field under the new gauge group $U(1)_X$ while $\Zp$ 
represents the new gauge boson.

The  scalar potential, with the usual SM Higgs doublet 
$H$, and two new scalars, namely the EW singlet  $S_1$ and 
the real singlet $S_2$, is given by
\begin{align}
& V(H,S_1,S_2) = -\mu_H^2 (H^\dagger H) - \mu_{S_1}^2 (S_1^\dagger S_1)  - \mu_{S_2}^2 S_2^2  
\nonumber \\
&+ \lambda_H (H^\dagger H)^2  +  \lambda_{HS_1} (H^\dagger H)(S_1^\dagger S_1) + \lambda_{S_1} (S_1^\dagger S_1)^2 
\nonumber \\
&+  \lambda_{S_2} S_2^4  +  \lambda_{HS_2} (H^\dagger H) S_2^2 
+ \lambda_{S_1 S_2} (S_1^\dagger S_1) S_2^2 
\nonumber \\
&+ \sigma_{1} S_2^3 +\sigma_{2}(H^{\dagger} H )S_2 +\sigma_{3}(S_1^{\dagger} S_1 )S_2 
\label{eq:potential}
\end{align}
where the parameters $\mu_H, \mu_{S_1}, \mu_{S_2}, \sigma_1, \sigma_2$ and $\sigma_3$ 
have mass dimensions while $\lambda_H, \lambda_{HS_1}, \lambda_{S_1}, \lambda_{S_2}, 
\lambda_{HS_2}$ and $\lambda_{S_1S_2}$ are real dimensionless couplings. The EW symmetry is spontaneously broken when the neutral component of the Higgs doublet $H$ gets 
a VEV while the additional $U(1)$ symmetry gets broken through the 
VEV of $S_1$. Then, in the unitary gauge, we can write the $H, S_1$ and $S_2$
fields as
\begin{align}\label{eq:vevs}
H = \frac{1}{\sqrt{2}}\begin{pmatrix}     0 \\     v_h + {\cal H} \end{pmatrix} , \,\,
 S_1 = \frac{1}{\sqrt{2}} (v_1 + {\cal S}_1), \,\,
S_2 =v_2+{\cal S}_2
\end{align}
where $v_{h}, v_1$ and $v_{2}$ are VEV's of corresponding scalar fields while 
${\cal H}, {\cal S}_1$ and ${\cal S}_2$ are the physical scalars in the gauge basis. Note that
the terms in the above scalar potential with coefficients ($\lambda_{HS_1},\lambda_{HS_2}, \lambda_{S_1S_2}, \sigma_2$ and $\sigma_3$) lead to a mixing between the three physical neutral scalars 
in the gauge basis, which we then choose to call $h,h_s$ and $s$ in the mass basis, 
once the fields have acquired VEV. We discuss the minimization conditions on the scalar potential, 
including constraints on the various coupling parameters ($\mu_i,\lambda_i,\sigma_i$) and the 
corresponding mass matrix relevant for this work in the {\tt Appendix}.

After the neutral scalar fields have acquired  VEV's, the SM gauge bosons ($Z, W^{\pm}$)  get mass
through the symmetry breaking via $<H>\, = v_h/\sqrt{2} \sim v_{EW} $ and the $\Zp$ gets mass 
via $<S_1> \,= v_1/\sqrt{2}$.
We can also write a mass term for the vector-like quark,
\begin{align}\label{Dmass}
\mathcal{L}_{\textrm{mass}} &= M_{x} \,\, \overline{xq}_L \, xq_R.
\end{align}
Note that the new exotic vector-like quark $xq$ has color, hypercharge, and an 
extra $U(1)_X$ interaction, but no $SU(2)_L$ interaction. Since this new $xq$ quark
is vector-like with respect to both $U(1)_Y$ as well as $U(1)_X$, the
model is anomaly free.  Without any other interaction, the $xq$ quark
will be stable. As none of the SM particles are charged under the
new $U(1)_X$ symmetry, the new symmetry remains hidden from
the SM, provided the gauge-kinetic-mixing terms are strongly
suppressed. However, its gauge quantum numbers allow flavour changing
Yukawa interactions with the SM quarks via
the singlet Higgs boson $S_1$.
\begin{align} \label{eq:Yextra}
\mathcal{L}_{\textrm{Y}_{extra}} &= Y_{xq} \,\, \overline{xq}_L \, q_{iR} \,\,S_1 + h.c.
\end{align}
where $q_{iR}$ can be either the up-type or down-type quarks depending on the 
hypercharge we assign to $xq$ for the above Lagrangian to be hypercharge 
singlet. We consider only mixing with the third generation quarks such that
the hypercharge of both $xq_L$ and $xq_R$ must be equal to that of either $t_R$ or $b_R$.
This also requires that the $U(1)_X$ charge ($Y_{X}$) for the
exotic quark $xq$ must satisfy $Y_{X}=q^\prime$. Such a term in the
Lagrangian leads to mixing between the top (bottom) quark with the new exotic
vector-like quark $xq$, giving rise to EW decay modes for the heavy quark. 
In addition we can also write interaction terms for the new scalar $S_2$ with the 
$xq$ given by:
\begin{align}\label{eq:lag}
\begin{split}
\mathcal{L}= -f_{X} \,\,\overline{xq} \,\, xq \,\,S_2 \,\, .
\end{split}
\end{align}
Note that the vector-like quark gets a bare mass as well as a mass from its Yukawa interaction with the singlet Higgs $S_2$, once $S_2$ gets a VEV.
Note that using the above Lagrangian,  the mass eigenstates from the mixing matrix 
for the $q$ and $xq$ (where $q=t \,(b)$ and $xq=xt \,(xb)$) along with their left and right 
mixing angles ($\theta_L,\theta_R$) can be 
determined using bi-unitary transformations. 

Expressing the gauge eigenstates for the
mixing quarks as $q^0$ and $xq^0$, the mass matrix in the ($q^0,xq^0$)
basis is given by
\begin{align} \label{bDmass}
\mathcal{M} &= \left(\begin{matrix}
 y_q \, v_h/\sqrt{2} &  0 \\
 Y_{xq} \, v_1/\sqrt{2} &  M_{xq}
\end{matrix}
\right).
\end{align}
where $y_q$ is the usual Yukawa coupling of the SM quark with the Higgs doublet $H$
while $M_{xq}=M_{x} - f_X v_2$.
This matrix can be diagonalized with a bi-unitary transformation
$\mathcal{M}_{\text{diag}}=\mathcal{O}_L \mathcal{M} \mathcal{O}_R^\dag$, where $\mathcal{O}_L$ 
and $\mathcal{O}_R$ are unitary matrices which rotate the left-chiral and right-chiral gauge eigenstates
to the mass eigenstates respectively. The interaction of the physical mass eigenstates
($q,xq$) can then be obtained by writing the gauge basis states as
\begin{align} \label{Dstates}
q_i^0 &=  q_i \cos\theta_i + xq_i \sin\theta_i
, &
xq_i^0 &= - q_i \sin\theta_i + xq_i \cos\theta_i \,\, ,
\end{align}
while the rotation matrices $\mathcal{O}_i$ are given by
\begin{align} \label{Drot}
\begin{aligned}\mathcal{R}_i &=
\begin{pmatrix}
 \cos\theta_i &  \sin\theta_i \\
 -\sin\theta_i & \cos\theta_i
\end{pmatrix},&
&\textrm{where } i=L,R.
\end{aligned}
\end{align}
The corresponding mixing angles for the left- and
right-handed fields follow from diagonalizing the matrices
$\mathcal{M}\mathcal{M}^{\dag}$ and $\mathcal{M}^{\dag}\mathcal{M}$.

For our purposes, we can safely assume the mixings to be very small. 
However, it must be noted that such mixings although small would still ensure that 
the vector-like quarks decay to SM quarks and bosons, i.e. $xq \to q' W, \,\, q Z, \,\, q h$. 
As the mixing angles $\theta_L$ and $\theta_R$ are 
constrained\footnote{A detailed description on vector-like quarks and mixing can be found 
in Ref. \cite{Aguilar-Saavedra:2013qpa}} by observables involving  $t,b$ quarks, in interactions within 
the SM as well as the entries in the 
Cabibbo-Kobayashi-Maskawa (CKM) matrix, the small values would help in avoiding any such constraints
easily. We also note that the model has three neutral scalars which also mix 
when $H, \, S_1$ and $S_2$ acquire VEV's. We must make the 125 GeV 
Higgs \cite{Chatrchyan:2012xdj, Aad:2012tfa} to be SM like 
and therefore dominantly the doublet component which therefore is unaffected in its properties 
by the presence of the exotic quark and scalar singlets. We however can try and allow significant mixing 
amongst the singlet scalars (see appendix). For simplicity we shall restrict our choice of the 
parameter space in the model, such that the mixing angles remain small.  

A few comments are in order here:
\begin{itemize}
\item A quick look at the scalar potential (Eq. \ref{eq:potential})  tells us that the mixing between 
the singlets and the doublet is related once the minimization conditions are imposed, as discussed 
in the appendix. 
\item One must also note that when the real singlet $S_2$ does not get a VEV, it is not possible to   
make the singlet-doublet (${\cal H}-{\cal S}_2$) mixing vanishingly small while making the two 
singlets (${\cal S}_1-{\cal S}_2$) mix substantially, as the mixing terms in the off-diagonal entries in the 
mass-squared matrix (${\cal M}^2_{13}, {\cal M}^2_{23}$) in Eq. \ref{eqn:hs1s2masstwo} are of equal strength (by Eq. \ref{eq:min} as $v_2=0$ and $v_h \sim v_1$), 
written in the (${\cal H},{\cal S}_1,{\cal S}_2$) basis. Note that ${\cal H}-{\cal S}_1$
mixing is independent of this and can be made negligibly small.
\item One can therefore achieve almost minimal mixing of the doublet with singlets while large mixing 
between the two singlets, once $S_2$ gets a VEV.
\end{itemize}

\section{Analysis}
Thus in our framework, we consider the most simplified scenario where the $xq \equiv xt$ has the 
same hypercharge as $t_R$ and therefore mixes with the top-quark. Note that
although the mixing angles ($\theta_L,\theta_R$) can be arbitary and small, it ensures 
a mixing which shall make the $xt$ decay. Again, small mixings, if allowed in the scalar 
sector between $H,S_1$ and $S_2$ also ensures that $xt$ which had a dominant 
coupling with $S_2$ now also couples to the different scalar mass eigenstates ($h, s$ and $h_s$). 
Here the $h$ is identified to be the SM like Higgs boson. Therefore the vector-like 
quark (VLQ) can decay through several modes if kinematically allowed. 

We must point out that the new $U(1)$ gauge boson mass is given by 
$M_{\Zp} = g_X q^\prime v_1$ where $v_1$ is the VEV of $S_1$ that breaks $U(1)_X$ and 
$q^\prime$ is the $U(1)_X$ quantum number of $S_1$. Since 
this $\Zp$ only couples to top quarks, it is not possible to produce this directly 
at colliders and therefore existing bounds on such a $\Zp$ are very 
weak. 
The possible production channels for such a top-phillic $\Zp$ would be via associated production
with $t\bar{t}$, and $xt \,\, \overline{xt}$ or it can have loop-induced productions: 
\begin{align}
 p p \to t\bar{t} \, \Zp \, ; \,\, xt \,\, \overline{xt} \, \Zp \, ; \,\, \Zp + j \,\, (loop) 
\end{align}
Note that in the absence of 
any kinetic mixing of the new $U(1)$ with SM $Z$, the $\Zp$ will have a four-body decay 
\begin{align*}
 \Zp \to b W^+ \,\, \bar{b} W^-  
\end{align*}
when $ 2 m_b + 2 m_W < m_{\Zp} < m_t + m_b + m_W$, while $\Zp$ will have the three-body decays
\begin{align*}
 \Zp \to  t \, \bar{b} \, W^- \,\, , \,\, \bar{t} \, b \, W^+
\end{align*}
when $ m_t + m_b + m_W < m_{\Zp} < 2 m_t$.
A detailed phenomenological account of such a 
top-phillic $\Zp$ can be found in Ref.~\cite{Greiner:2014qna, Cox:2015afa}. Note that in our model, 
the $\Zp$ has an additional mode of production which may become significant for lighter VLQ masses
as well as the strength of the $U(1)_X$ gauge coupling, $g_X$. 
So the $\Zp$ can be much lighter than the heavier scalar mass eigenstates $s$ and $h_s$ as well 
as the VLQ. Thus $xt$ can have quite a few possible decay products through the channels:
\begin{align}
xt \to bW^+, t h, t Z, t s, t h_s, t \Zp
\label{eq:xtdks}
\end{align}
provided the mass states of $s,\, h_s, \, \Zp$ are lighter than $xt$. The additional decay modes 
would lead to new signals for the VLQ which can be produced at the LHC through strong interactions. 
As the existing bounds on such VLQ rely on its decay via $bW^+, \,\, t h, \,\, t Z$ modes 
only \cite{Aad:2015tba, Khachatryan:2015oba}, 
the additional decay modes are expected to dilute the existing bounds on their mass and therefore one 
can have significantly lighter top-like VLQ still allowed by the experimental data. Signals for a VLQ with 
new decay modes to light neutral scalar has been considered before, for e.g. in Ref.~\cite{Karabacak:2014nca}.
However as we want to scan over the VLQ mass to fit the diphoton excess, there would be regions 
of parameter space where the VLQ becomes lighter than some of the above mentioned states, 
namely $h_s, s$ or $\Zp$ which would disallow its decay to them.  Since we set the mass of $s$ to 
be 750 GeV, lighter $xq$ can still decay through the remaining channels listed in Eq. \ref{eq:xtdks}. 
A much detail analysis of the VLQ and $\Zp$ signal at LHC in this model, which is interesting in itself is 
planned as future work.

\begin{figure}[t]
\includegraphics[width=3.1in,height=1.1in]{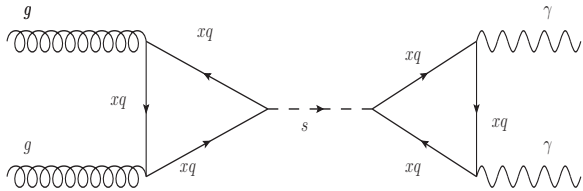}
\caption{Feynman diagram representing the diphoton resonant production through the scalar $s$ at 
LHC.}
\label{fig:feyn}
\end{figure}

For our current analysis, we shall consider the spectrum where $s$ is dominantly composed of 
$S_2$ with a mass of around 750 GeV. Just like the VLQ, the scalar $s$ can also decay via new 
channels other than a pair of SM particles including the Higgs boson ($h$). Namely, the new modes 
can be summarized as 
\begin{align}
s \to \Zp\Zp, h_s h_s, xt \,\, \bar{t},t \,\, \overline{xt}, h \,\, h_s \,\, , 
\end{align}
again the decay being possible, depending on the mass of the decay products. 
The important thing to note here is that $s$ would decay to gluon-pair as well as diphoton via 
one-loop diagrams very similar to the SM Higgs boson, with the dominant contribution coming from 
$xt$ in the loop (since $S_2$ couples to $xt$ directly with a coupling strength $f_X$, which can be large).
Thus the production of this 750 GeV scalar is determined by the mass of $xt$ and the size of the 
coupling strength $f_X$. The other decay channels for $s$ can help in increasing the decay width of the 
resonance.\footnote{A wider resonance can also be realised ($\sim 45$ GeV) with both physical 
singlet masses $(m_{s}, m_{h_s})$ being close to around $\sim 750$ GeV and separated by a small
mass splitting. As both can contribute to the diphoton final state signal, it would be possible to explain 
the large width observed by the experimental collaborations without each scalar resonance being 
very wide itself.} The loop induced decay of the $s$ to the massless gauge boson pairs $gg$ is given 
by the effective Lagrangian
\begin{align} \label{eq:sgg}
\mathcal{L}_{sGG} = - \lambda_{sgg} s ~G_{\mu\nu} G^{\mu\nu} 
\end{align}
where $\lambda_{sgg} = \alpha_s f_X F_{1/2} (\tau_{xq}) /( 16 \pi M_{xq}) $. We choose 
the definition of the loop-function $F_{1/2} (\tau_{xq})$ as given in Ref.~\cite{Djouadi:2005gi}. 
We neglect the mixing effects here which can be justified by assuming that they are small 
enough to be negligible for the production rates but relevant to ensure the new decay modes for $s$ 
and $xq$. However, as the new decay modes can reduce the branching fractions of the 
$s \to \gamma\gamma$, in order to fit the excess data, the mixings would be constrained to a 
great extent. For example,
\begin{itemize}
 \item As the $s \to \Zp \Zp, h \, h_s, h_s \, h_s$ decays happen when ${\cal S}_1-{\cal S}_2$ mix, this mixing has to be taken small when the above decays are kinematically allowed for lighter 
 $\Zp, h_s$ so as not to suppress the diphoton mode significantly. In the current analysis we shall 
 assume this mixing to be suppressed.  Note that $s \to h h$ is disallowed 
 by our choice of negligible mixing of the singlet scalars with the doublet Higgs as discussed in the 
 appendix. 
\end{itemize}
We use the above interaction to calculate the production rates for the 
scalar $s$ at the LHC run II and show the dependence of the cross section on the 
mass $M_{xq}$ normalized to the coupling $f_X$. To do this we have implemented the effective vertex 
given by Eq. \ref{eq:sgg} in the event generator {\tt CalcHEP} \cite{Belyaev:2012qa} and also include 
running of the strong coupling constant $\alpha_s$ calculated at $m_s=750$ GeV in our estimates. 
The rates for the process shown in Fig. \ref{fig:feyn} is then simply given by the product of the 
production rate multiplied to the diphoton branching fraction which is naively 
$\alpha_{em}^2/\alpha_s^2(m_s) \lesssim 0.7\%$ at most if no additional decay modes of $s$ are present.%

\begin{figure}[hb!]
\includegraphics[width=3.2in,height=2.6in]{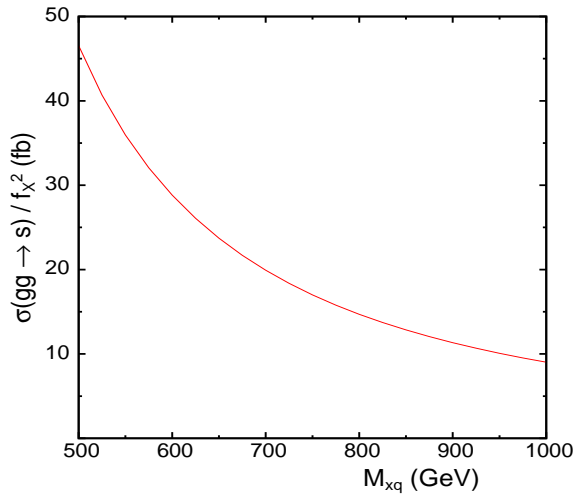}
\caption{The on-shell production cross section of $s$ through gluon-gluon fusion at LHC with 
$\sqrt{s}=13$ TeV as a function of the VLQ mass ($M_{xq}$). Note that we have normalized the
cross section with the coupling strength squared ($f_X^2$).}
\label{fig:sprod}
\end{figure}

In Fig. \ref{fig:sprod} we plot the leading-order (LO) production cross section for $s$ with 
mass $m_s = 750$ GeV through the gluon-gluon fusion at the LHC with $\sqrt{s}=13$ TeV as a 
function of the VLQ mass ($M_{xq}$). 
The cross sections are shown to be normalized with the coupling strength squared given by $f_X^2$.
We find that with only a single VLQ and without including any QCD corrections to the production, 
the production is as large as 46 fb for $M_{xq}=500$ GeV and drops to about 10 fb when 
$M_{xq}=1$ TeV with $f_X=1$. We also find that with $xt$, the branching fraction for 
$s \to \gamma\gamma$ is about 0.6\% which falls dramatically down to 0.04\% if the VLQ is $xb$ 
(due to the additional suppression from electric charge $(Q_d^2/Q_u^2)^2 \equiv 1/16$ to the 
partial width) for all values of $M_{xq}$. Note that the production cross section for the $s$ is 
independent of this choice and therefore, quite clearly $xt$  helps in enhancing the diphoton 
rates compared to $xb$. Assuming that $f_X \simeq \sqrt{4\pi}$ is taken at its perturbative limit, 
the production rates for $s$ are enhanced 
by a factor of $\sim 12.57$ which means that a resonant diphoton cross section with the top-like 
VLQ can be $\simeq$10 fb with $M_{xt} \simeq 375$ GeV while achieving it with the bottom type VLQ 
will be clearly impossible. Of course one must note here that the QCD $K$-factors for the $gg \to s$
production should not be very different from that of the SM Higgs. Including the QCD 
corrections can therefore simply double the production cross section ($K_{NNLO} \simeq 2 $), 
thus pushing the upper limit on $M_{xq}$ to about 450 GeV to get $\sim 10$ fb diphoton rate.
For values of the VLQ mass less than $m_s/2$, the tree-level decay 
mode, $s \to \overline{xq} \,\, xq$ opens up. This would completely dominate over all other decay 
channels making it practically impossible to fit the diphoton excess. Thus $M_{xq} > m_s/2$ 
provides a lower bound to our choice of the VLQ mass. Quite clearly, one must resort to 
non-perturbative coupling strengths $f_X$ for heavier VLQ mass to get the required cross section 
in the diphoton channel when including a single VLQ in the particle spectrum.  

We however must point out that adding more generations of $xq$ can easily alleviate this tension on the 
mass of the VLQ and the coupling $f_X$. Working within a single generation of VLQ, one can also 
include the bottom like partner ($xb$) with the same $U(1)_X$ charge as the top partner ($xt$) as the 
minimal matter content in the model. This actually helps in increasing the production cross section 
by a factor of 4, assuming $M_{xb}=M_{xt}$ and $f_{xb}=f_{xt}$. However, the $s \to \gamma\gamma$ 
branching in this case drops to 0.25\% which still effectively gives an enhancement of about 5/3 to the 
diphoton rates. This rate can  be further enhanced by adding much lighter and less constrained 
vector-like charged leptons ($x\tau$) that could enhance the photon branching significantly, thus easing 
the tension on the VLQ masses. In fact we find that for a single $x\tau$ with mass of about 400 GeV, the 
$s \to \gamma \gamma$ branching fraction peaks and goes up by a factor of $\sim 9$ to about 4.5\%, 
provided $f_{x\tau} \sim 3$, while $f_{xq}=1$ and $M_{xq}=600$ GeV. 
This would satisfy the 10 fb limit for diphoton cross section with just $xt$ as the VLQ with 
$M_{xt} \simeq 775 (1050)$ GeV without (with) $K$-factors, thus easily meeting the current limits 
on VLQ mass. Notably, adding more vector-like particles charged under the $U(1)_X$ gauge 
symmetry also enriches the $\Zp$ phenomenology of the model with additional production and 
decay channels. We leave these interesting possibilities to be taken up in a future work.  

\begin{figure}[hb!]
\includegraphics[width=2.8in,height=2.8in]{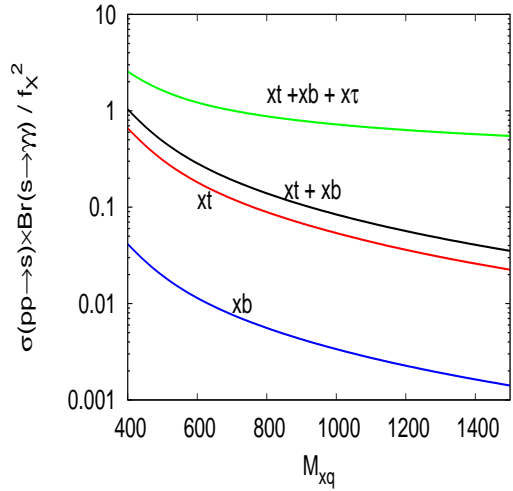}
\caption{The diphoton production rate for different exotic fermion scenarions with $\sqrt{s}=13$ TeV as 
a function of the VLQ mass ($M_{xq}$). $M_{x\tau} = 400$ GeV has been taken.}
\label{fig:sigmabr}
\end{figure}
To show the relative dependence of including different set of vector-like fermions in the 
particle spectrum, we plot the LO cross section of the diphoton signal at LHC with $\sqrt{s}=13$ TeV
in Fig. \ref{fig:sigmabr} as a function of the vector-like fermion mass while keeping their 
Yukawa couplings with the singlet $s$ to be equal ($f_X=1$), which essentially represents the 
normalization condition ($1/f_{X}^2$) that we have put for the cross section. The $M_{xq}$ has been 
varied between $400-1500$ GeV. We can see that with just $xb$, it would be hard to achieve the 
observed diphoton signal. However, in all other scenarios, it is quite easy to observe a signal of $1-10$ fb 
for the diphoton rate with perturbative values for $f_X$. As observed
earlier, the inclusion of a vector-like charged lepton with $M_{x\tau}=400$ GeV allows the 
required signal rate, where VLQ masses are as high as 1.5 TeV. 
Note that a wide range of coupling and mass for the VLQ can easily accommodate the observed 
resonant signal. To summarize the plots, we note that the limit is around 700 GeV in 
the only top-like case. If both top-like and bottom-like VLQ are included, a little higher values in the 
mass  of VLQ i.e. around 900 GeV is achievable as the $gg\rightarrow s$ production cross-section 
becomes four times that of only top-like VLQ case. By including a 400 GeV vector-like charged lepton, 
it is possibble to push the vector-like quark masses above 1 TeV. Independent of the exotic fermion 
content, $f_X$ values below 0.5 is not allowed as long as the limit for di-photon production rate 
is between $1-10$ fb (at LO).

Thus we find that within our model framework and a minimal extension of the matter particles 
by a single generation we can easily accommodate the diphoton excess without reverting to 
non-perturbative couplings or a very high new physics scale. However, as already mentioned earlier, 
in our model we have new decay modes (Eq. \ref{eq:xtdks}) for the  VLQ that not only relaxes the 
current limits on their masses but also leads to interesting signatures at the LHC which we leave for 
future analyses.  We also expect that with more data collected by the experiment, the dijet resonance 
may show up at the same invariant mass for which the diphoton signal 
is observed (since the branching of $s\to gg$ can be significantly large for most of the parameter 
space), unless the other aforementioned decay modes of $s$ become large. 
In addition, a very interesting signature in the model could be the production of 
light $\Zp$ through decays of the primarily produced VLQ's. 
\section{Summary}  
In this work we show that a simple $U(1)$ extension to the SM gauge symmetry with a minimal set of 
new particles can easily accommodate the excess without invoking a large enough scale for new physics
or non-perturbative couplings. We argue that with a high new physics scale, explaining the 
diphoton excess may lead to large non-perturbative coupling strengths for particle interactions.  
We show that a required low scale can be very easily realised within the context of our 
``hidden symmetry" model, thus keeping all couplings perturbative as well as complying with 
experimental constraints on the new physics scale. 

We show that in our model the observed diphoton excess also highlights some new interesting signals that 
should show up as more data is accumulated in the run II of LHC. 
We perform a simplistic  scan on the relevant parameters to show the compatibility of the 
resonant diphoton data with our model predictions. We also highlight the possibility of a very light 
$\Zp$ in the model with sub-TeV mass that can appear in decay cascades of the heavier particles such
as the VLQ produced at the LHC. We leave the phenomenological analysis of such possibilities in 
our model as future work.  
\begin{widetext}
\section{Appendix}
\label{sec:appendix}
We discuss the scalar potential of our model in some detail here. To find the minimum of the 
potential we use the following extremization conditions given by $\frac{\partial V}{\partial {\cal H}} = 0$, 
$\frac{\partial V}{\partial {\cal S}_1} = 0$ and $ \frac{\partial V}{\partial {\cal S}_2} = 0$ which give us the following 
equations respectively:

 \begin{align} \label{eq:min}
 \lambda_H  v_h^3 + \frac{1}{2} \lambda_{HS_1}  v_1^2 v_h 
 + \lambda_{HS_2} v_2^2v_h + \sigma_2 v_2 v_h  - \mu_H^2 v_h  & =  0 \,\, , \nonumber \\ 
  \lambda_{S_1} v_1^3 + \frac{1}{2}  \lambda_{HS_1}  v_1 v_h^2
 + \lambda_{S_1S_2} v_1 v_2^2   + \sigma_3 v_1v_2 - \mu_{S_1}^2 v_1  & =  0   \,\, , \\ \nonumber
 \lambda_{HS_2}  v_2 v_h^2 + \lambda_{S_1S_2}  v_1^2 v_2 + 4 \lambda_{S_2} v_2^3 
 + 3 \sigma_1 v_2^2  + \frac{1}{2}( \sigma_2 v_h^2 + \sigma_3 v_1^2 ) - 2 \mu_{S_2}^2 v_2  & =  0   \,\, .
 \end{align}
 Note that for the potential to be bounded from below we have 
 \begin{align}
   \lambda_H > 0, && \lambda_{S_1} > 0, && \lambda_{S_2} > 0.
 \end{align} 
Using Eq.\ref{eq:min} we can substitute for $\mu_H, \mu_{S_1}$ and $\mu_{S_2}$ in the scalar potential. 
Then the mass square matrix for the three neutral scalars in the gauge basis 
(${\cal H},{\cal S}_1,{\cal S}_2$) becomes
\begin{equation}\label{eqn:hs1s2masstwo}
\mathcal{M}^2 =\begin{pmatrix}   2 \lambda_H v_h^2 & \lambda_{HS_1} v_1v_h & (\sigma_2 
+ 2 \lambda_{HS_2} v_2) v_h \\
\lambda_{HS_1} v_1 v_h   & 2 \lambda_{S_1}  v_1^2 & (\sigma_3 + 2 \lambda_{S_1S_2} v_2) v_1 \\
(\sigma_2 + 2 \lambda_{HS_2} v_2 ) v_h & (\sigma_3 + 2 \lambda_{S_1S_2} v_2 ) v_1 & 
\frac{1}{2v_2} (2 (8 \lambda_{S_2} v_2  + 3 \sigma_1) v_2^2 - \sigma_2 v_h^2 - \sigma_3 v_1^2)
\end{pmatrix}  \,\, .
\end{equation}
 For the point $({\cal H}=0, {\cal S}_1=0, {\cal S}_2 =0 )$ to be a minima of the potential, the matrix 
 $\mathcal{M}^2$ should be positive definite, which is possible if its $3$ upper left determinants are 
 positive. The corresponding conditions are given below
\begin{align}
  2 \lambda_H v_h^2 > 0 \,\, ;  &&
\begin{vmatrix}   2 \lambda_H v_h^2 & \lambda_{HS_1} v_1v_h \\
\lambda_{HS_1}  v_1v_h & 2 \lambda_{S_1} v_1^2
\end{vmatrix} 
 > 0 \implies 4  \lambda_H\lambda_{S_1} - \lambda_{HS_1}^2 > 0  \,\, ; &&
 |\mathcal{M}^2| > 0   \,\, .
 \end{align}
For simplicity we have assumed that the mixing of ${\cal H}$ with ${\cal S}_1$ and ${\cal S}_2$ 
is vanishingly small and we shall set it to be zero. Note that such a choice not only imposes the 
condition that the scalar ${\cal H}\equiv h$ is a pure doublet component but also that it will have the 
exact properties of the SM Higgs boson with mass $m_h=\sqrt{2 \lambda_H} v_h \simeq 125$ GeV.  
A quick look at the mass matrix then gives the conditions $\lambda_{HS_1}=0$ and 
$\sigma_2+2 \lambda_{HS_2} v_2=0$ for non-zero $v_h$ and $v_1$.

We can now consider the two remaining singlet scalars independent of the doublet-component 
$\cal H$. The reduced mass square matrix for the ${\cal S}_1$ and ${\cal S}_2$ sector becomes
\begin{equation}\label{eqn:s1s2mass}
M =\begin{pmatrix}  
2 \lambda_{S_1}  v_1^2 & (\sigma_3 + 2 \lambda_{S_1S_2} v_2) v_1 \\
 (\sigma_3 + 2 \lambda_{S_1S_2} v_2) v_1  & \frac{1}{2v_2} (2 (8 \lambda_{S_2} v_2  + 3 \sigma_1) v_2^2 
 - \sigma_2 v_h^2 - \sigma_3 v_1^2)
\end{pmatrix}  \,\, .
\end{equation}
%
The fields $({\cal S}_1,{\cal S}_2)$ can now be expressed in terms of mass eigenstates $(h_s,\,\,s)$ where 
\begin{eqnarray}
{\cal S}_1 = h_s \cos{\beta} - s \sin{\beta}   \,\, , \\
{\cal S}_2 = h_s \sin{\beta} + s \cos{\beta} ~.
\end{eqnarray}
The mixing angle $\beta$ is given by 
\begin{equation}
 \tan{2\beta} = \frac{2M_{12}}{M_{11}-M_{22}}
\end{equation}
and
\begin{equation}
 \sin{2\beta} = \frac{2M_{12}}{\sqrt{(M_{11}-M_{22})^2+4M_{12}^2}}  \,\, ,
\end{equation}
where ${M}_{ij}$ is the $(i,j)^{th}$ element of ${M}$ in Eq.\ref{eqn:s1s2mass}.

The mass eigenvalues for the two scalars $s$ and $h_s$ are then given by
\begin{equation}
 m_1^2 = \frac{1}{2}\Big(M_{11}+M_{22}-\sqrt{(M_{11}-M_{22})^2+4M_{12}^2}\Big)
\end{equation}
and 
\begin{equation}
 m_{2}^2 =  \frac{1}{2}\Big(M_{11}+M_{22}+\sqrt{(M_{11}-M_{22})^2+4M_{12}^2}\Big)  \,\, .
\end{equation}

%
%
For our analysis we have $m_h=125$ GeV, $m_s=750$ GeV while $m_{h_s}$ is a free parameter 
which we can vary in the model. Note that it is possible to make $h_s$ lighter than $s$ as well 
the vector-like quarks by choosing parameters such that $M_{11} < M_{22}$. In the absence of 
mixing with the Higgs doublet, the $h_s$ then decays to SM quarks through mixing of the VLQ
with SM quarks.

Note that while the condition for the non-mixing of the doublet with either of the singlets may not 
forbid the couplings of the singlet $s$ with $h$, but it shall prevent the decay of $s$ to any SM 
particle pair arising out of such mixings in the scalar sector. In fact the condition  
$\sigma_2+2 \lambda_{HS_2} v_2=0$ leads to the exact cancellation 
of an interaction vertex between $h-h-s$  arising from the terms in the scalar potential given 
by $ + \lambda_{HS_2} (H^\dagger H) S_2^2  +\sigma_{2}(H^{\dagger} H )S_2 $. This is crucial 
in avoiding the possible decay of the 750 GeV singlet scalar to SM Higgs pair which is constrained by 
data \cite{Franceschini:2015kwy}. Similarly, the decay of $h_s$ to $h$ pair is also forbidden 
due to the mixing suppression.
The relevant vertices for the interactions within the scalar sector can be easily 
determined for the mass eigenstates and are given by \\[2mm]
\begin{center}
\begin{tabular}{lcl} 
${h}_{}$ \phantom{-} ${h_s}_{}$ \phantom{-} ${h_s}_{}$ \phantom{-}  &:&
	$-2  \lambda_{HS_2} v_h  \, \, s_{\beta}{}^2  $\\[2mm]
${h}_{}$ \phantom{-} ${h_s}_{}$ \phantom{-} $s{}_{}$ \phantom{-}  &:&
	$-2  \lambda_{HS_2}  v_h \, \, c_{\beta}  s_{\beta} $\\[2mm]
${h}_{}$ \phantom{-} $s{}_{}$ \phantom{-} $\,\,\, s{}_{}$ \phantom{-}  &:&
	$-2  \lambda_{HS_2}  v_h \,\,c_{\beta}{}^2 $\\[2mm]
${h_s}_{}$ \phantom{-} ${h_s}_{}$ \phantom{-} $s{}_{}$ \phantom{-}  &:&
	$\big(6 c_{\beta}{}^2  s_{\beta} \lambda_{S_1} v_1-24 c_{\beta} s_{\beta}{}^2  \lambda_{S_2} v_2-2 (2-3 s_{\beta} {}^2) s_{\beta} \lambda_{S_1S_2} v_1$ \\[2mm]
  &:& $-2 (1-3 s_{\beta} {}^2) c_{\beta} \lambda_{S_1S_2} v_2-6 c_{\beta} s_{\beta}{}^2  \sigma_1- (1-3 s_{\beta} {}^2) c_{\beta} \sigma_3\big)$\\[2mm]
${h_s}_{}$ \phantom{-} $s{}_{}$ \phantom{-} $\,\,\, s{}_{}$ \phantom{-}  &:&
	$-\big(6 c_{\beta} s_{\beta}{}^2  \lambda_{S_1} v_1+24 c_{\beta}{}^2  s_{\beta} \lambda_{S_2} v_2+2 (1-3 s_{\beta} {}^2) c_{\beta} \lambda_{S_1S_2} v_1$ \\[2mm]
  &:& $-2 (2-3 s_{\beta} {}^2) s_{\beta} \lambda_{S_1S_2} v_2+6 c_{\beta}{}^2  s_{\beta} \sigma_1- (2-3 s_{\beta} {}^2) s_{\beta} \sigma_3\big)$
\\ 
\end{tabular}
\end{center}
where $c_\beta=\cos\beta$ and $s_\beta=\sin\beta$.

A few benchmark points can be identified which give possibilities of a spectrum where $m_s \sim 750$
GeV while $m_{h_s}$ is either heavier, lighter or has mass close to $s$. For example:
\begin{align*}
 (\sigma_1,\sigma_3) = (-150, 65) ~\rm{GeV}, && 
 (  \lambda_{S_1} ,\lambda_{S_2}, \lambda_{S_1S_2}, \lambda_{HS_2})  = (1, 0.2,-0.04, 0.05)  \,\, , &&   
 (v_h, v_1,v_2) = (246, 750, 760) ~\rm{GeV}   ,
\end{align*}
gives $m_{h_s}=1.06$ TeV while $m_s = 749.1$ GeV with a very 
small mixing ($|\sin\beta| \sim 5.6 \times 10^{-3}$). Similarly,
\begin{align*}
 (\sigma_1,\sigma_3) = (-150, 65) ~\rm{GeV}, && 
 (  \lambda_{S_1} ,\lambda_{S_2}, \lambda_{S_1S_2}, \lambda_{HS_2})  = (1, 0.2,-0.05, 0.05)  \,\, , &&   
 (v_h, v_1,v_2) = (246, 450, 750) ~\rm{GeV},
\end{align*}
gives $m_{h_s}=636.4$ GeV while $m_s = 746.2$ GeV with again a 
suppressed mixing angle ($|\sin\beta| \sim 1.5 \times 10^{-2}$). However a slight variation 
in the model parameters also gives for
\begin{align*}
 (\sigma_1,\sigma_3) = (-130, 90) ~\rm{GeV}, && 
 (  \lambda_{S_1} ,\lambda_{S_2}, \lambda_{S_1S_2}, \lambda_{HS_2})  = (1, 0.19,-0.05, 0.1)  \,\, , &&   
 (v_h, v_1,v_2) = (246, 531, 760) ~\rm{GeV}, 
\end{align*}
$m_{h_s}=758.7$ GeV while $m_s = 747.8$ GeV with  a not so
suppressed mixing angle ($|\sin\beta| \sim 0.54$) which can give the possibility of two 
resonances look like a single wide resonance, as observed by the ATLAS collaboration.





\end{widetext}
\begin{acknowledgments}
\emph {Acknowledgments:} 
S.K.R. thanks T. Li for fruitful discussions. This work was partially supported by funding available 
from the Department of Atomic Energy, Government of India, for the Regional Centre for 
Accelerator-based Particle Physics (RECAPP), Harish-Chandra Research Institute.
\end{acknowledgments}



\begin{thebibliography}{99}
  
\bibitem{lhc1}
  The ATLAS collaboration,
  ATLAS-CONF-2015-081;
 \bibitem{lhc2} 
   The CMS collaboration,
  CMS PAS EXO-15-004. 


\bibitem{Harigaya:2015ezk} 
  K.~Harigaya and Y.~Nomura,
  arXiv:1512.04850 [hep-ph].

\bibitem{Mambrini:2015wyu} 
  Y.~Mambrini, G.~Arcadi and A.~Djouadi,
  arXiv:1512.04913 [hep-ph].

\bibitem{Backovic:2015fnp} 
  M.~Backovic, A.~Mariotti and D.~Redigolo,
  arXiv:1512.04917 [hep-ph].

\bibitem{Angelescu:2015uiz} 
  A.~Angelescu, A.~Djouadi and G.~Moreau,
  arXiv:1512.04921 [hep-ph].
  
\bibitem{Nakai:2015ptz} 
  Y.~Nakai, R.~Sato and K.~Tobioka,
  arXiv:1512.04924 [hep-ph].
  
\bibitem{Knapen:2015dap} 
  S.~Knapen, T.~Melia, M.~Papucci and K.~Zurek,
  arXiv:1512.04928 [hep-ph].
  
\bibitem{Buttazzo:2015txu} 
  D.~Buttazzo, A.~Greljo and D.~Marzocca,
  arXiv:1512.04929 [hep-ph].
  
\bibitem{Pilaftsis:2015ycr} 
  A.~Pilaftsis,
  arXiv:1512.04931 [hep-ph].

\bibitem{Franceschini:2015kwy} 
  R.~Franceschini {\it et al.},
  arXiv:1512.04933 [hep-ph].

\bibitem{DiChiara:2015vdm} 
  S.~Di Chiara, L.~Marzola and M.~Raidal,
  arXiv:1512.04939 [hep-ph].
  
\bibitem{Higaki:2015jag} 
  T.~Higaki, K.~S.~Jeong, N.~Kitajima and F.~Takahashi,
  arXiv:1512.05295 [hep-ph].

\bibitem{McDermott:2015sck} 
  S.~D.~McDermott, P.~Meade and H.~Ramani,
  arXiv:1512.05326 [hep-ph].

\bibitem{Ellis:2015oso} 
  J.~Ellis, S.~A.~R.~Ellis, J.~Quevillon, V.~Sanz and T.~You,
  arXiv:1512.05327 [hep-ph].

\bibitem{Low:2015qep} 
  M.~Low, A.~Tesi and L.~T.~Wang,
  arXiv:1512.05328 [hep-ph].

\bibitem{Bellazzini:2015nxw} 
  B.~Bellazzini, R.~Franceschini, F.~Sala and J.~Serra,
  arXiv:1512.05330 [hep-ph].

\bibitem{Gupta:2015zzs} 
  R.~S.~Gupta, S.~Jäger, Y.~Kats, G.~Perez and E.~Stamou,
  arXiv:1512.05332 [hep-ph].

\bibitem{Petersson:2015mkr} 
  C.~Petersson and R.~Torre,
  arXiv:1512.05333 [hep-ph].

\bibitem{Molinaro:2015cwg} 
  E.~Molinaro, F.~Sannino and N.~Vignaroli,
  arXiv:1512.05334 [hep-ph].


\bibitem{Dutta:2015wqh} 
  B.~Dutta, Y.~Gao, T.~Ghosh, I.~Gogoladze and T.~Li,
  arXiv:1512.05439 [hep-ph].

\bibitem{Cao:2015pto} 
  Q.~H.~Cao, Y.~Liu, K.~P.~Xie, B.~Yan and D.~M.~Zhang,
  arXiv:1512.05542 [hep-ph].

\bibitem{Kobakhidze:2015ldh} 
  A.~Kobakhidze, F.~Wang, L.~Wu, J.~M.~Yang and M.~Zhang,
  arXiv:1512.05585 [hep-ph].

\bibitem{Cox:2015ckc} 
  P.~Cox, A.~D.~Medina, T.~S.~Ray and A.~Spray,
  arXiv:1512.05618 [hep-ph].

\bibitem{Ahmed:2015uqt} 
  A.~Ahmed, B.~M.~Dillon, B.~Grzadkowski, J.~F.~Gunion and Y.~Jiang,
  arXiv:1512.05771 [hep-ph].

\bibitem{Becirevic:2015fmu} 
  D.~Becirevic, E.~Bertuzzo, O.~Sumensari and R.~Z.~Funchal,
  arXiv:1512.05623 [hep-ph].

\bibitem{No:2015bsn} 
  J.~M.~No, V.~Sanz and J.~Setford,
  arXiv:1512.05700 [hep-ph].

\bibitem{Demidov:2015zqn} 
  S.~V.~Demidov and D.~S.~Gorbunov,
  arXiv:1512.05723 [hep-ph].

\bibitem{Chao:2015ttq} 
  W.~Chao, R.~Huo and J.~H.~Yu,
  arXiv:1512.05738 [hep-ph].

\bibitem{Fichet:2015vvy} 
  S.~Fichet, G.~von Gersdorff and C.~Royon,
  arXiv:1512.05751 [hep-ph].

\bibitem{Curtin:2015jcv} 
  D.~Curtin and C.~B.~Verhaaren,
  arXiv:1512.05753 [hep-ph].

\bibitem{Bian:2015kjt} 
  L.~Bian, N.~Chen, D.~Liu and J.~Shu,
  arXiv:1512.05759 [hep-ph].

\bibitem{Chakrabortty:2015hff} 
  J.~Chakrabortty, A.~Choudhury, P.~Ghosh, S.~Mondal and T.~Srivastava,
  arXiv:1512.05767 [hep-ph].

\bibitem{Agrawal:2015dbf} 
  P.~Agrawal, J.~Fan, B.~Heidenreich, M.~Reece and M.~Strassler,
  arXiv:1512.05775 [hep-ph].


\bibitem{Csaki:2015vek} 
  C.~Csaki, J.~Hubisz and J.~Terning,
  arXiv:1512.05776 [hep-ph].

\bibitem{Falkowski:2015swt} 
  A.~Falkowski, O.~Slone and T.~Volansky,
  arXiv:1512.05777 [hep-ph].

\bibitem{Bai:2015nbs} 
  Y.~Bai, J.~Berger and R.~Lu,
  arXiv:1512.05779 [hep-ph].

\bibitem{Benbrik:2015fyz} 
  R.~Benbrik, C.~H.~Chen and T.~Nomura,
  arXiv:1512.06028 [hep-ph].

\bibitem{Kim:2015ron} 
  J.~S.~Kim, J.~Reuter, K.~Rolbiecki and R.~R.~de Austri,
  arXiv:1512.06083 [hep-ph].

\bibitem{Gabrielli:2015dhk} 
  E.~Gabrielli, K.~Kannike, B.~Mele, M.~Raidal, C.~Spethmann and H.~Veermäe,
  arXiv:1512.05961 [hep-ph].

\bibitem{Alves:2015jgx} 
  A.~Alves, A.~G.~Dias and K.~Sinha,
  arXiv:1512.06091 [hep-ph].

\bibitem{Megias:2015ory} 
  E.~Megias, O.~Pujolas and M.~Quiros,
  arXiv:1512.06106 [hep-ph].

\bibitem{Carpenter:2015ucu} 
  L.~M.~Carpenter, R.~Colburn and J.~Goodman,
  arXiv:1512.06107 [hep-ph].

\bibitem{Bernon:2015abk} 
  J.~Bernon and C.~Smith,
  arXiv:1512.06113 [hep-ph].

\bibitem{Han:2015cty} 
  C.~Han, H.~M.~Lee, M.~Park and V.~Sanz,
  arXiv:1512.06376 [hep-ph].

\bibitem{Chang:2015bzc} 
  S.~Chang,
  arXiv:1512.06426 [hep-ph].

\bibitem{Han:2015dlp} 
  H.~Han, S.~Wang and S.~Zheng,
  arXiv:1512.06562 [hep-ph].

\bibitem{Luo:2015yio} 
  M.~x.~Luo, K.~Wang, T.~Xu, L.~Zhang and G.~Zhu,
  arXiv:1512.06670 [hep-ph].

\bibitem{Chang:2015sdy} 
  J.~Chang, K.~Cheung and C.~T.~Lu,
  arXiv:1512.06671 [hep-ph].

\bibitem{Bardhan:2015hcr} 
  D.~Bardhan, D.~Bhatia, A.~Chakraborty, U.~Maitra, S.~Raychaudhuri and T.~Samui,
  arXiv:1512.06674 [hep-ph].

\bibitem{Feng:2015wil} 
  T.~F.~Feng, X.~Q.~Li, H.~B.~Zhang and S.~M.~Zhao,
  arXiv:1512.06696 [hep-ph].

\bibitem{Liao:2015tow} 
  W.~Liao and H.~q.~Zheng,
  arXiv:1512.06741 [hep-ph].

\bibitem{Cho:2015nxy} 
  W.~S.~Cho, D.~Kim, K.~Kong, S.~H.~Lim, K.~T.~Matchev, J.~C.~Park and M.~Park,
  arXiv:1512.06824 [hep-ph].

\bibitem{Barducci:2015gtd} 
  D.~Barducci, A.~Goudelis, S.~Kulkarni and D.~Sengupta,
  arXiv:1512.06842 [hep-ph].

\bibitem{Chao:2015nsm} 
  W.~Chao,
  arXiv:1512.06297 [hep-ph].

\bibitem{Ding:2015rxx} 
  R.~Ding, L.~Huang, T.~Li and B.~Zhu,
  arXiv:1512.06560 [hep-ph].

\bibitem{Han:2015qqj} 
  X.~F.~Han and L.~Wang,
  arXiv:1512.06587 [hep-ph].

\bibitem{Antipin:2015kgh} 
  O.~Antipin, M.~Mojaza and F.~Sannino,
  arXiv:1512.06708 [hep-ph].

\bibitem{Wang:2015kuj} 
  F.~Wang, L.~Wu, J.~M.~Yang and M.~Zhang,
  arXiv:1512.06715 [hep-ph].

\bibitem{Cao:2015twy} 
  J.~Cao, C.~Han, L.~Shang, W.~Su, J.~M.~Yang and Y.~Zhang,
  arXiv:1512.06728 [hep-ph].

\bibitem{Huang:2015evq} 
  F.~P.~Huang, C.~S.~Li, Z.~L.~Liu and Y.~Wang,
  arXiv:1512.06732 [hep-ph].

\bibitem{Heckman:2015kqk} 
  J.~J.~Heckman,
  arXiv:1512.06773 [hep-ph].

\bibitem{Bi:2015uqd} 
  X.~J.~Bi, Q.~F.~Xiang, P.~F.~Yin and Z.~H.~Yu,
  arXiv:1512.06787 [hep-ph].

\bibitem{Kim:2015ksf} 
  J.~S.~Kim, K.~Rolbiecki and R.~R.~de Austri,
  arXiv:1512.06797 [hep-ph].

\bibitem{Berthier:2015vbb} 
  L.~Berthier, J.~M.~Cline, W.~Shepherd and M.~Trott,
  arXiv:1512.06799 [hep-ph].

\bibitem{Cline:2015msi} 
  J.~M.~Cline and Z.~Liu,
  arXiv:1512.06827 [hep-ph].

\bibitem{Bauer:2015boy} 
  M.~Bauer and M.~Neubert,
  arXiv:1512.06828 [hep-ph].

\bibitem{Chala:2015cev} 
  M.~Chala, M.~Duerr, F.~Kahlhoefer and K.~Schmidt-Hoberg,
  arXiv:1512.06833 [hep-ph].

\bibitem{Dev:2015isx} 
  P.~S.~B.~Dev and D.~Teresi,
  arXiv:1512.07243 [hep-ph].

\bibitem{deBlas:2015hlv} 
  J.~de Blas, J.~Santiago and R.~Vega-Morales,
  arXiv:1512.07229 [hep-ph].

\bibitem{Boucenna:2015pav} 
  S.~M.~Boucenna, S.~Morisi and A.~Vicente,
  arXiv:1512.06878 [hep-ph].

\bibitem{Murphy:2015kag} 
  C.~W.~Murphy,
  arXiv:1512.06976 [hep-ph].

\bibitem{Hernandez:2015ywg} 
  A.~E.~C.~Hernández and I.~Nisandzic,
  arXiv:1512.07165 [hep-ph].

\bibitem{Dey:2015bur} 
  U.~K.~Dey, S.~Mohanty and G.~Tomar,
  arXiv:1512.07212 [hep-ph].

\bibitem{Pelaggi:2015knk} 
  G.~M.~Pelaggi, A.~Strumia and E.~Vigiani,
  arXiv:1512.07225 [hep-ph].

\bibitem{Belyaev:2015hgo} 
  A.~Belyaev, G.~Cacciapaglia, H.~Cai, T.~Flacke, A.~Parolini and H.~Serôdio,
  arXiv:1512.07242 [hep-ph].
  
\bibitem{huang:2015} 
  W.~C.~Huang, Y.~L.~S.~Tsai and T.~C.~Yuan,
  arXiv:1512.07268 [hep-ph].
 
\bibitem{moretti:2015} 
  S.~Moretti and K.~Yagyu,
  arXiv:1512.07462 [hep-ph].

\bibitem{patel:2015} 
  K.~M.~Patel and P.~Sharma,
  arXiv:1512.07468 [hep-ph].

\bibitem{badziak:2015} 
  M.~Badziak,
  arXiv:1512.07497 [hep-ph].
  
\bibitem{raychaudhuri:2015} 
  S.~Chakraborty, A.~Chakraborty and S.~Raychaudhuri,
  arXiv:1512.07527 [hep-ph].
  
\bibitem{Cao:2015} 
  Q.~H.~Cao, S.~L.~Chen and P.~H.~Gu,
  arXiv:1512.07541 [hep-ph].

\bibitem{Zupan:2015} 
  W.~Altmannshofer, J.~Galloway, S.~Gori, A.~L.~Kagan, A.~Martin and J.~Zupan,
  arXiv:1512.07616 [hep-ph].

\bibitem{Cvetic:2015chl} 
  M.~Cvetič, J.~Halverson and P.~Langacker,
  arXiv:1512.07622 [hep-ph].
  
\bibitem{Gu:2015gl} 
  J.~Gu and Z.~Liu,
  arXiv:1512.07624 [hep-ph].
   
\bibitem{Chakraborty:2015jvs} 
  I.~Chakraborty and A.~Kundu,
  arXiv:1512.06508 [hep-ph].
 
\bibitem{Grossmann:2010wm} 
  B.~N.~Grossmann, B.~McElrath, S.~Nandi and S.~K.~Rai,
  Phys.\ Rev.\ D {\bf 82}, 055021 (2010)
  doi:10.1103/PhysRevD.82.055021
  [arXiv:1006.5019 [hep-ph]].


\bibitem{Karabacak:2014nca} 
  D.~Karabacak, S.~Nandi and S.~K.~Rai,
  Phys.\ Lett.\ B {\bf 737}, 341 (2014)
  doi:10.1016/j.physletb.2014.08.065
  [arXiv:1405.0476 [hep-ph]].

\bibitem{Aguilar-Saavedra:2013qpa} 
  J.~A.~Aguilar-Saavedra, R.~Benbrik, S.~Heinemeyer and M.~Pérez-Victoria,
  Phys.\ Rev.\ D {\bf 88}, no. 9, 094010 (2013)
  doi:10.1103/PhysRevD.88.094010
  [arXiv:1306.0572 [hep-ph]].

\bibitem{Chatrchyan:2012xdj} 
  S.~Chatrchyan {\it et al.} [CMS Collaboration],
  Phys.\ Lett.\ B {\bf 716}, 30 (2012)
  doi:10.1016/j.physletb.2012.08.021
  [arXiv:1207.7235 [hep-ex]].
  
\bibitem{Aad:2012tfa} 
  G.~Aad {\it et al.} [ATLAS Collaboration],
  Phys.\ Lett.\ B {\bf 716}, 1 (2012)
  doi:10.1016/j.physletb.2012.08.020
  [arXiv:1207.7214 [hep-ex]].

\bibitem{Greiner:2014qna} 
  N.~Greiner, K.~Kong, J.~C.~Park, S.~C.~Park and J.~C.~Winter,
  JHEP {\bf 1504}, 029 (2015)
  doi:10.1007/JHEP04(2015)029
  [arXiv:1410.6099 [hep-ph]].
  
\bibitem{Cox:2015afa} 
  P.~Cox, A.~D.~Medina, T.~S.~Ray and A.~Spray,
  arXiv:1512.00471 [hep-ph]. 

\bibitem{Aad:2015tba} 
  G.~Aad {\it et al.} [ATLAS Collaboration],
  Phys.\ Rev.\ D {\bf 92}, no. 11, 112007 (2015)
  doi:10.1103/PhysRevD.92.112007
  [arXiv:1509.04261 [hep-ex]].
  
\bibitem{Khachatryan:2015oba} 
  V.~Khachatryan {\it et al.} [CMS Collaboration],
  Phys.\ Rev.\ D {\bf 93}, no. 1, 012003 (2016)
  doi:10.1103/PhysRevD.93.012003
  [arXiv:1509.04177 [hep-ex]].

\bibitem{Djouadi:2005gi} 
  A.~Djouadi,
  Phys.\ Rept.\  {\bf 457}, 1 (2008)
  doi:10.1016/j.physrep.2007.10.004
  [hep-ph/0503172].

\bibitem{Belyaev:2012qa} 
  A.~Belyaev, N.~D.~Christensen and A.~Pukhov,
  Comput.\ Phys.\ Commun.\  {\bf 184}, 1729 (2013)
  [arXiv:1207.6082 [hep-ph]].

\end{thebibliography}
\end{document}